\documentclass[prd,preprint,groupedaddress,onecolumn,showpacs,10pt]{revtex4-1}
\usepackage{hyperref}
\usepackage{graphicx,xcolor}
\usepackage{xtab,afterpage}
\usepackage{lipsum}
\usepackage{bm}
\usepackage{amssymb,amsmath,amsfonts, mathrsfs}

\newcommand{\be}{\begin{equation}}
\newcommand{\ee}{\end{equation}}
\newcommand{\ba}{\begin{array}}
\newcommand{\ea}{\end{array}}
\newcommand{\bqa}{\begin{eqnarray}}
\newcommand{\eqa}{\end{eqnarray}}

\usepackage{indentfirst}
\usepackage{booktabs}
\usepackage{tabularx}
\usepackage{makecell}
\usepackage{float}
\usepackage{appendix}
\usepackage{graphicx}
\usepackage{epsfig}
\usepackage{epstopdf}
\usepackage{amsmath, amssymb}
\usepackage{makecell}
\usepackage{multirow}
\usepackage{booktabs}
\usepackage{makecell}
\usepackage{color}
\usepackage{appendix}
\usepackage{graphicx}
\usepackage{epsfig}
\usepackage{epstopdf}
\usepackage{amsmath, amssymb}
\usepackage{multirow}
\usepackage{hyperref}
\usepackage{longtable}

\begin{document}


\title{Updated analysis of  charmonium states in a relativized quark potential model }


\author{Xiu-Li Gao}
\affiliation{School of Physics, Southeast University, Nanjing 211189, P.~R.~China}
\author{Jun-Xi Cui}
\affiliation{School of Physics, Southeast University, Nanjing 211189, P.~R.~China}
\author{Yu-Hui Zhou}
\affiliation{School of Physics, Southeast University, Nanjing 211189, P.~R.~China}
\author{ Zhi-Yong Zhou}
\email[]{zhouzhy@seu.edu.cn}
\affiliation{School of Physics, Southeast University, Nanjing 211189, P.~R.~China}
\date{\today}


\begin{abstract}
Motivated by recent experimental observations of charmonium(-like) states, we investigate their interpretation as conventional charmonium states within the framework of relativized quark potential model. We found a consistent description of the updated masses of charmonia, bottomonia and some selected charmed mesons, then systematically compute  open-charm strong decay widths and electromagnetic transition widths for radially excited charmonium states up to $n=5$. The numerical results show significantly improved agreement with experimental masses and widths. Notably, the newly-observed $\chi_{c1}(4010)$ and $\chi_{c1}(4274)$ are tentatively associated  with the $\chi_{c1}(2P)$ and $\chi_{c1}(3P)$ respectively, while the $\chi_{c0}(4500)$ and $\chi_{c0}(4700)$ are interpreted as the $\chi_{c0}(4P)$ and $\chi_{c0}(5P)$ states. For the  controversial $\chi_{c1}(3872)$ state, we suggest  its potential connection to $\chi_{c1}(4010)$, as a two-pole structure due to the $\chi_{c1}(2P)$ coupling to $D\bar{D}$ channels. Intriguingly, the predicted mass of $\chi_{c0}(2P)$ is about 3851 MeV, diverging from the original estimate of $3916$ MeV in the Godfrey and Isgur's work  but consistent with non-relativistic potential model predictions. This calculation might provide more insight to the future experimental investigation of charmonium(-like) states.
\end{abstract}


\maketitle

\section{Introduction}\label{section1}
Since the discovery of the $J/\psi$ meson in 1974, over thirty charmonium(-like) states have been identified through experimental analyses. Because of a significant increase in the number of observed states, the charmonium spectrum appears denser than the predictions made by the conventional quark potential model~\cite{ParticleDataGroup:2024cfk}. To comprehend the richness of the meson spectrum, many theoretical efforts focus on elucidating the mechanisms behind the additional states observed experimentally. Many approaches are devoted to explore mechanisms beyond the $q\bar{q}$ paradigm, such as multiquark states, hadronic molecules, hybrid states, and other  exotic degrees of freedom, to account for the surplus on charmonium(-like) spectrum~(see literatures in refs.~\cite{Guo:2017jvc,Chen:2016qju}). There is also some other attempts aiming to refine the potential model to accommodate a more densely structured spectrum~\cite{Li:2009ad,Deng:2016stx}.

Puzzles surrounding the charmonium spectrum can be exemplified by  the $J^{PC} =1^{++}$ states. The $\chi_{c1}(3872)$~\cite{Belle:2003nnu}, identified as the first excited $1^{++}$ charmonium state, has sparked intense debate since its discovery. However, positioned near the $D_0\bar D_0^*$ threshold, the $\chi_{c1}(3872)$ can not be easily categorized as $\chi_{c1}(2P)$ state. Firstly, its mass notably deviates from predictions in relativistic Godfrey-Isgur~(GI) models~\cite{Godfrey:1985xj} and  nonrelativistic potential~(NR) models~\cite{Barnes:2005pb}. Furthermore, its decay width falls significantly below the anticipated value for $\chi_{c1}(2P)$ in different decay models. The higher state $\chi_{c1}(4140)$ \cite{CDF:2009jgo} also exhibit a mass and a width far exceeding expectations of $\chi_{c1}(2P)$. More and more researches suggest an exotic nature of the $\chi_{c1}(3872)$, while the whereabouts of the $\chi_{c1}(2P)$ state remains an open query.  A recent breakthrough by the LHCb collaboration revealed the presence of a new state, $\chi_{c1}(4010)$, in the amplitude analysis of $B^+\rightarrow D^{*\pm}D^\mp K^+$ decays~\cite{LHCb:2024vfz}. This discovery seems to bridge the long-standing gap of the $1^{++}$ charmonium spectroscopy, but more evidences from other experiments are required to confirm this discovery.
Similar challenges are encountered in categorizing  $0^{++}$  states. There also exist long-time debates about $\chi_{c0}(2P)$ state and its potential correlation  with the reported $\chi_{c0}(3860)$~\cite{Belle:2017egg} or $\chi_{c0}(3915)$ states~\cite{Belle:2004lle,BaBar:2007vxr}. The conclusion has not been made until now. For the higher states, recent observations by LHCb of  the $0^{++}$  $\chi_{c0}(4500)$ and $\chi_{c0}(4700)$~\cite{LHCb:2016nsl,LHCb:2016axx}, which are also suggested to possess exotic origins, as reviewed in the PDG review~\cite{ParticleDataGroup:2024cfk}. These observed states have  further enriched  discussions about how to understand the charmonium-like spectrum.

It is agreed that QCD permits the formation of exotic meson states beyond conventional  $q\bar{q}$ configurations~\cite{Weinberg:2013cfa}.  However, at present it is still absent for a unified theoretical framework which could simultaneously describe conventional  $q\bar{q}$ and exotic states. Thus, identifying conventional $q\bar{q}$ candidates provides a critical baseline for distinguishing non-$q\bar{q}$ configurations. The GI model has long served as a reference for such studies. Yet accumulating evidence demonstrates that its predictive power diminishes for states  above the open-flavor threshold, where significant discrepancies between its predictions and observed masses emerge.  Despite these shortcomings, the GI framework retains conceptual validity for understanding meson spectroscopy. In our opinion, one possibility of its weak prediction power above the open-flavor threshold is that it is too ambitious to describe the states containing quarks from the light to the heavy. If it is limited in studying a subset of the mesonic states, the predictive scope of this model might be extended after recalibrating its parameters.

Motivated by the advances in experimental investigation, we revisit charmonium spectroscopy within the framework of relativistic quark potential model, and systematically evaluate open-charm decay properties and electromagnetic transitions.
While this study focuses on charmonium states, we incorporate bottomonium states and ground-state $c\bar{u}$/$c\bar{s}$ systems to constrain model parameters through comparative analysis.
With the spectrum and wavefunctions extracted from the relativized framework, open-charm strong decay widths for states with radial quantum numbers up to $n=5$ are computed within the Quark Pair Creation (QPC) model~\cite{PhysRevD.53.3700}, and then compared with experimental measurements. Furthermore, we calculate electromagnetic transition rates, including E1 and M1 partial widths, which serve as references for future experimental explorations.
Our findings demonstrate that this updated implementation reconciles several longstanding discrepancies in charmonium decay patterns, and offer a more coherent spectroscopic interpretation consistent with contemporary data.

The paper is organized as follows. Theoretical background is presented  in Sec.~\ref{section2}, in which we briefly review the relativistic quark model  in Sec.~\ref{seciton21} and the QPC model in Sec.~\ref{section22}, and then illustrate how to compute the radiative transitions in Sec.~\ref{section23}. The numerical results and subsequent discussions are devoted in Sec.~\ref{section3}.  The last section is the summary and conclusions.

\section{Theoretical Background}\label{section2}

\subsection{Brief review about the relativized quark potential model}\label{seciton21}

First, we briefly review the calculation scheme of the relativistic GI model~\cite{Godfrey:1985xj}.
The Hamiltonian of a system containing  a  quark and an anti-quark can be expressed as the sum of the relativistic kinematic term and the interaction potential
\begin{align}
H=\sqrt{\textbf{p}^2+m_1^2}+\sqrt{\textbf{p}^2+m_2^2}+V(\textbf{p},\textbf{r}),
\end{align}
where $m_1$ and $m_2$ represent the masses of constituent quarks and $\textbf{p}$ denote the c.m. momentum.
In the non-relativistic limit, the momentum-dependent potential, defined as $V(\textbf{p},\textbf{r})=H^{conf}+H^{hyp}+H^{so}$,
includes the linear confinement and color-Coulomb interactions
\bqa
H^{conf}=-\left(\frac{3}{4}c+\frac{3}{4}br-\frac{\alpha_s(r)}{r}\right),
\eqa
the color-hyperfine interaction
\bqa
H^{hyp}=-\frac{\alpha_s(r)}{m_1m_2}\left[\frac{8\pi}{3}\mathbf{S_1}\cdot\mathbf{S_2}\delta^3(\mathbf{r})+\frac{1}{r^3}(\frac{3\mathbf{S_1}\cdot\mathbf{r}\mathbf{S_2}\cdot\mathbf{r}}{r^2}-\mathbf{S_1}\cdot\mathbf{S_2})\right],
\eqa
and the spin-orbit interaction
\bqa
H^{so}=-\frac{\alpha_s(r)}{r^3}\left(\frac{1}{m_1}+\frac{1}{m_2}\right)\left(\frac{\mathbf{S_1}}{m_1}+\frac{\mathbf{S_2}}{m_2}\right)\cdot\mathbf{L}-\frac{1}{2r}\frac{\partial H^{conf}}{\partial r}\left(\frac{\mathbf{S_1}}{m_1^2}+\frac{\mathbf{S_2}}{m_2^2}\right)\cdot \mathbf{L}.
\eqa

The explicit treatment of relativistic corrections addresses the non-negligible velocity-dependent effects, particularly manifest in their hyperfine splitting. The ``relativization" procedure starts from parameterizing the running coupling constant w.r.t the transition momentum $Q$ as $\alpha(Q^2)=\sum_k\alpha_k e^{-Q^2/4\gamma_k^2}=0.25e^{-Q^2}+0.15e^{-Q^2/10}+0.20e^{-Q^2/1000}$.
To regularize the long-range interaction behavior, a smearing function $\rho(\mathbf{r}-\mathbf{r'})=\frac{\sigma^3}{\pi^{3/2}}e^{-\sigma^2(\mathbf{r'}-\mathbf{r})2}$ is introduced to find the smearing potentials through
\bqa
\tilde{V}(r)=\int\mathrm{d}^3\mathbf{r'}\rho(\mathbf{r}-\mathbf{r'})V(r')
\eqa
where $\sigma$ is a smearing parameter related to the quark flavors.

The smeared color Coulomb potential between quark $i$ and quark $j$ in the configuration space  is defined by Fourier Transformation as
\bqa
\tilde G(r)=-\sum_k\frac{4\alpha_k}{3r}(\frac{2}{\sqrt{\pi}}\int_0^{\tau_{kij}r}e^{-x^2\mathrm{d}x}),
\eqa
where $\tau_{kij}^2=(\gamma_k^{-2}+\sigma_{ij}^{-2})^{-1}$ and $\sigma_{ij}^2=\sigma_0^2\left[\frac{1}{2}+\frac{1}{2}\left(\frac{4m_im_j}{(m_i+m_j)^2}\right)^4\right]+s^2\left(\frac{2m_im_j}{m_i+m_j}\right)^2$.
Similarly, the linear confinement potential in the configuration space could be transformed to a smeared form as
\bqa
\tilde S(r)=br[\frac{1}{\sqrt{\pi}\sigma_{ij}r}e^{-\sigma_{ij}^2r^2}+(1+\frac{1}{2\sigma_{ij}^2r^2})\frac{2}{\sqrt{\pi}}\int_0^{\sigma_{ij}r}e^{-x^2}\mathrm{d}x]+c.
\eqa
With the help of these two important ingredients, the interaction potential is expressed as the sum of the six effective  terms as
\begin{align}
V(\textbf{p},\textbf{r})= G_\text{eff}^{Coul}(r) +G_\text{eff}^{so(v)}(r)+G_\text{eff}^{con}(r)+G_\text{eff}^{ten}(r)+\tilde S(r)+S_\text{eff}^{so(s)}(r),
\end{align}
and the terms of the right side are  respectively the effective color Coulomb,  spin-orbit, contact, tensor, linear confinement and scalar spin-orbit interaction potential. These terms are written down as
\begin{align}\label{eq:modpotential}
G_\text{eff}^{coul}(r)=&\left(1+\frac{p^2}{E_1 E_2}\right)^{1 / 2} \tilde G(r)\left(1+\frac{p^2}{E_1 E_2}\right)^{1 / 2},\nonumber\\
G_\text{eff}^{so(v)}(r)=&\left(\frac{\mathbf{S}_1 \cdot \mathbf{L}}{2 m_1^2} \frac{1}{r} \frac{\partial \tilde G_{11}^{\mathrm{so}(v)}}{\partial r}+\frac{\mathbf{S}_2 \cdot \mathbf{L}}{2 m_2^2} \frac{1}{r} \frac{\partial \tilde G_{22}^{\mathrm{so}(v)}}{\partial r}+\frac{\left(\mathbf{S}_1+\mathbf{S}_2\right) \cdot \mathbf{L}}{m_1 m_2} \frac{1}{r} \frac{\partial \tilde G_{12}^{\mathrm{so}(v)}}{\partial r}\right) ,\nonumber\\
G_\text{eff}^{con}(r)=&\frac{2 \mathbf{S}_1 \cdot \mathbf{S}_2}{3 m_1 m_2} \nabla^2 \tilde G_{12}^c,\nonumber\\
G_\text{eff}^{ten}(r)=&-\left(\frac{\mathbf{S}_1 \cdot \widehat{r} \mathbf{S}_2 \cdot \widehat{r}-\frac{1}{3} \mathbf{S}_1 \cdot \mathbf{S}_2}{m_1 m_2}\right)\left(\frac{\partial^2}{\partial r^2}-\frac{1}{r} \frac{\partial}{\partial r}\right) \tilde G_{12}^t, \nonumber\\
S_\text{eff}^{so(s)}(r)=&-\frac{\mathbf{S}_1 \cdot \mathbf{L}}{2 m_1^2} \frac{1}{r} \frac{\partial \widetilde{S}_{11}^{\mathrm{so}(s)}}{\partial r}-\frac{\mathbf{S}_2 \cdot \mathbf{L}}{2 m_2^2} \frac{1}{r} \frac{\partial \widetilde{S}_{22}^{\mathrm{so}(s)}}{\partial r}.
\end{align}
The superscripts and subscripts of the $\tilde G$ and $\tilde S$ terms in eq.~(\ref{eq:modpotential}) stand for explicit treatment of relativistic corrections, where the modifications could be generally represented as
\bqa
\tilde G_{ij}^\alpha=(\frac{m_i m_j}{E_i E_j})^{1/2+\epsilon_\alpha}\tilde G(r)(\frac{m_i m_j}{E_i E_j})^{1/2+\epsilon_\alpha},
\eqa
and $\alpha=c,\ t,\ so(v),\ so(s)$ represents contact, tensor, vector spin-orbit, and scalar spin-orbit interactions respectively.

The eigenfunction of Hamiltonian is solved through diagonalization within an extensive basis of simple harmonic oscillator~(SHO) wave functions. This computational approach leverages the unique analytic structure of SHO wavefunctions, whose isomorphic function forms in momentum space, $\psi_{nlm}(p,\theta,\varphi)$, and coordinate space,  $\phi_{nlm}(r,\theta,\varphi)$, are given by similar Hermite polynomials multiplied by Gaussian factors. Such duality allows the separation of momentum-dependent and coordinate-dependent components in the interaction matrix elements, enabling computational efficient evaluation through factorized integrals in a simple way. That means, the meson wave functions in the momentum space and the configuration space are expanded in series of  the SHO wave functions as \cite{Blundell:1996as},
\begin{align}
\psi_{nlm}(p,\theta,\varphi)&=\sum_{n'}f_{n'}R_{{n'}l}(p)Y_{lm}(\Omega_p),\\
\phi_{nlm}(r,\theta,\varphi)&=\sum_{n'}f_{n'}\tilde{R}_{{n'}l}(r)Y_{lm}(\Omega_r),
\end{align}
where $f_{n'}$ is the coefficient of the SHO wave function basis, and the radial parts  in momentum space $R_{nl}$ and that in configuration space $\tilde{R}_{nl}$ can be respectively expressed as
\begin{align}
R_{nl}(p)&=\frac{(-1)^n(-i)^l}{\beta^{\frac{3}{2}}}\sqrt{\frac{2^{l+2-n}(2l+2n+1)!!}{\sqrt{\pi}n![(2l+1)!!]^2}}\left(\frac{p}{\beta }\right)^le^{-\frac{1}{2}\frac{p^2}{\beta^2}}{_1\textit{F}_1}\left(-n,l+\frac{3}{2},\frac{r^2}{\beta^2}\right),\\
\tilde{R}_{nl}(r)&=\beta^{\frac{3}{2}}\sqrt{\frac{2^{l+2-n}(2l+2n+1)!!}{\sqrt{\pi}n![(2l+1)!!]^2}}(\beta r)^le^{-\frac{1}{2}\beta^2r^2}{_1\textit{F}_1}\left(-n,l+\frac{3}{2},\beta^2r^2\right),
\end{align}
where $\beta$ is a harmonic oscillator parameter in wave function and ${_1\textit{F}_1}$ the confluent hypergeometric function.

When we calculate the Hamiltonian matrix element of the operator with both the momentum and the spacial coordinate as $f_1(p)f_2(r)$, a complete set of the wave function is inserted so that the element could be represented as the matrix product of two terms as
\begin{align}
\langle i|f_1(p)f_2(r)|j\rangle=\sum_n\langle i|f_1(p)|n\rangle\langle n|f_2(r)|j\rangle.
\end{align}
The matrix elements $\langle i|f_1(p)|n\rangle$ could be calculated in momentum space while the matrix elements $\langle n|f_2(r)|j\rangle$  are calculated in configuration space. The similarity of the SHO wave function in the momentum and configuration space makes numerical calculation very simple. Finally, we could find the wave functions of the eigenstates by diagonizing Hamiltonian matrix
in a large SHO basis. In the practical procedure, we find that numerical convergence is achieved by using 30 SHO basis states with the harmonic oscillator parameter $\beta=0.4$~GeV.


\subsection{The quark pair creation model} \label{section22}

The QPC model, alternatively known as the $^3P_0$ model due to its creation of $J^{PC}=0^{++}$ quark-antiquark pairs from the vacuum~\cite{MICU1969521,PhysRevD.8.2223,PhysRevD.9.1415,PhysRevD.53.3700}, is employed to calculate the open-charm decay widths of charmonium states.
This model has been demonstrated to provide reliable predictions for Okubo-Zweig-Iizuka (OZI) allowed strong decays into two-body hadronic final states, which dominate the decay patterns of conventional quarkonium states above the open-flavor threshold~\cite{PhysRevLett.104.122001,PhysRevD.69.054008,PhysRevD.81.094003,SEGOVIA2012322}.  The interaction could be derived from the quantum field theory Hamiltonian
\bqa
H_I=\gamma \int\mathrm{d}^3x\bar\psi(x)\psi(x),\ \ \  t=0,
\eqa
where $\psi(x)$ is a Dirac field operator associated with the space-time point $x$.
One could find, for the process of $A\rightarrow BC$, the $S$-matrix is expressed as
\begin{align}
\langle BC|S|A\rangle = I-2 \pi i \delta\left(E_f-E_i\right) \langle BC|T|A\rangle,
\end{align}
where the transition operator $T$  can be expressed as
\begin{align}
T=-3 \gamma \sum_m\langle 1 m; 1-m \mid 00\rangle \int \mathrm{d}^3 \vec{p}_3 \mathrm{~d}^3 \vec{p}_4 \delta^3\left(\vec{p}_3+\vec{p}_4\right) \mathcal{Y}_{1m}\left(\frac{\vec{p}_3-\vec{p}_4}{2}\right) \chi_{1\ -m}^{34} \phi_0^{34} \omega_0^{34} b_3^{\dagger}\left(p_3\right) d_4^{\dagger}\left(p_4\right).\label{Tmatrix}
\end{align}
The $\gamma$ parameter represents the production strength of a quark-antiquark pair from the vacuum. The superscripts and subscripts ``3" and ``4" represent the quark and antiquark. $\chi_{1-m}^{34}$, $\phi_0^{34}$, $\omega_0^{34}$ are spin, flavor and color wave functions of the quark pair generated from the vacuum, and the solid harmonics $\mathcal{Y}_1^m(\vec{p})=pY_1^m(\Omega_{p})$ represents its relative space wave function. If the mock state  of  meson $A$ is defined as
\begin{align}
\left|A\left(n_A{ }^{2 S_A+1} L_{A J_A M_{J_A}}\right)\left(\vec{P}_A\right)\right\rangle \equiv & \sqrt{2 E_A} \sum_{M_{L_A}, M_{S_A}}\left\langle L_A M_{L_A} S_A M_{S_A} \mid J_A M_{J_A}\right\rangle \notag\\
& \times \int \mathrm{d}^3 \vec{p}_A \psi_{n_A L_A M_{L_A}}\left(\vec{p}_A\right) \chi_{{S_A}M_{S_A}}^{12} \phi_A^{12} \omega_A^{12} \notag\\
& \times\left|q_1\left(\frac{m_1}{m_1+m_2} \vec{P}_A+\vec{p}_A\right) \bar{q}_2\left(\frac{m_2}{m_1+m_2} \vec{P}_A-\vec{p}_A\right)\right\rangle,\label{mockstate}
\end{align}
the matrix element of $T$ operator in the c.m. frame is
\begin{align}
\langle BC|T|A\rangle=&\delta^3\left(\vec{P}_B+\vec{P}_C-\vec{P}_A\right) \mathcal{M}^{M_{J_A} M_{J_B} M_{J_C}}\notag\\
= & \gamma \sqrt{8 E_A E_B E_C} \sum_{M_{L_A}, M_{S_A}, M_{L_B}, M_{S_B},M_{L_C}, M_{S_C},m}\left\langle L_A M_{L_A} S_A M_{S_A} \mid J_A M_{J_A}\right\rangle \notag\\
& \times\left\langle L_B M_{L_B} S_B M_{S_B} \mid J_B M_{J_B}\right\rangle\left\langle L_C M_{L_C} S_C M_{S_C} \mid J_C M_{J_C}\right\rangle \notag\\
& \times\langle 1 m 1-m \mid 00\rangle\left\langle\chi_{S_B M_{S_B}}^{14} \chi_{S_C M_{S_C}}^{32} \mid \chi_{S_A M_{S_A}}^{12} \chi_{1-m}^{34}\right\rangle \notag\\
& \times\left\langle\phi_B^{14} \phi_C^{32} \mid \phi_A^{12} \phi_0^{34}\right\rangle I\left(\vec{P}, m_1, m_2, m_3\right),
\end{align}
where $\left\langle\chi_{S_B M_{S_B}}^{14}\chi_{S_C M_{S_C}}^{32} \mid \chi_{S_A M_{S_A}}^{12} \chi_{1-m}^{34}\right\rangle$ is the spin matrix element, the color matrix element $\left\langle\omega_B^{14} \omega_C^{32} \mid \omega_A^{12} \omega_0^{34}\right\rangle=1/3$, and the flavor matrix element $\left\langle\phi_B^{14} \phi_C^{32} \mid \phi_A^{12} \phi_0^{34}\right\rangle=1/\sqrt{3}$ in this study. The space integral factor $I(\vec{P}, m_1, m_2,m_3)$ is given by
\begin{align}
I\left(\vec{P}, m_1, m_2, m_3\right)= & \int \mathrm{d}^3 \vec{p} ~\psi_{n_B L_B M_{L_B}}^*\left(\frac{m_3}{m_1+m_3} \vec{P}+\vec{p}\right) \psi_{n_C L_C M_{L_C}}^*\left(\frac{m_3}{m_2+m_3} \vec{P}+\vec{p}\right) \notag\\
& \times \psi_{n_A L_A M_{L_A}}(\vec{P}+\vec{p}) \mathcal{Y}_{1m}(\vec{p}),\label{amplitudeM}
\end{align}
where $\vec{P}=\vec{P_B}=-\vec{P_C}$ representing the three momentum of $B$ in the c.m. frame of $BC$ system.

By choosing the momentum $\vec{P}$ along the $z$-axis and  applying the Jacob-Wick formula \cite{JACOB2000774}, the partial wave amplitude is expressed by the helicity amplitude  $\mathcal{M}^{M_{J_A} M_{J_B} M_{J_C}}$ as
\begin{align}
 \mathcal{M}^{J L}(P)=&\frac{\sqrt{4 \pi(2 L+1)}}{2 J_A+1} \times \sum_{M_{J_B}, M_{J_C}}\left\langle L 0 JM_{J_A} \mid J_AM_{J_A}\right\rangle \notag\\
& \times\left\langle J_B M_{J_B} J_C M_{J_C} \mid J_AM_{J_A}\right\rangle \times \mathcal{M}^{M_{J_A} M_{J_B} M_{J_C}}(P \hat{z}).
\end{align}
Thus,  the partial decay width of meson $A$ could be calculated  through the following formula
\begin{align}
\Gamma=\frac{\pi}{4} \frac{P}{M_A^2}\sum_{S, L}\left|M^{S L}\right|^2.
\end{align}

    \subsection{Radiative transition model} \label{section23}

The quark-photon interaction in the tree-level QED framework is governed by the Hamiltonian density
\begin{align}
H_{em}=-\sum_{j}e_j\bar{\psi_j}\gamma_\mu^jA^{\mu}(\mathbf{k},\mathbf{r})\psi_j,
\end{align}
where $\psi_j$ denotes the Dirac field operator for the $j$-th constituent quark with electric charge $e_j$ and $\mathbf{k}$  the three momentum of photon.  $A^{\mu}(\mathbf{k},\mathbf{r})=\epsilon^\mu e^{i\mathbf{k}\cdot\mathbf{r}_j}$
represents the photon' quantized electromaganetic field in Coulomb gauge with polarization vector $\epsilon^\mu $.

Applying the Foldy-Wouthuysen transformation to derive the nonrelativistic reduction, the interaction Hamiltonian $H_{em}$ reduces to the leading-order multipole expansion as \cite{CLOSE1970477,PhysRevD.50.5639,PhysRevC.56.1099}
\begin{align}
h_e\simeq\sum_j\left[e_j\mathbf{r}_j\cdot\bm{\epsilon}-\frac{e_j}{2m_j}\bm{\sigma}_j(\bm{\epsilon}\times\hat{\mathbf{k}})\right]e^{-i\mathbf{k\cdot r}_j},
\end{align}
where $m_j$, $\mathbf{r}_j$ and $\mathbf{\sigma}_j$ are the mass, coordinate and Pauli matrix of the $j$-th quark, respectively. The $\mathbf{\epsilon}$ denotes the transverse polarization vector with $\mathbf{\epsilon}\cdot\mathbf{k}=0$. The first and second terms generate electric dipole~(E1) and magnetic dipole~(M1) transitions respectively.

The helicity amplitude of the electric and magnetic transitions between the initial state $J\lambda$ and final state $J'\lambda'$ can be expressed as
\begin{align}
\mathcal{A}_{\lambda}=-i\sqrt{\frac{\omega_\gamma}{2}}\langle J'\lambda'|h_e|J\lambda\rangle,
\end{align}
where
\begin{align}
\mathcal{A}_\lambda^E &=-i\sqrt{\frac{\omega_\gamma}{2}}\langle J'\lambda'|\sum_fe_j\mathbf{r}_j\cdot\bm{\epsilon}e^{-i\mathbf{k\cdot r}_j}|J\lambda\rangle,\\
\mathcal{A}_\lambda^M &=+i\sqrt{\frac{\omega_\gamma}{2}}\langle J'\lambda'|\sum_f\frac{e_j}{2m_j}\bm{\sigma}_j(\bm{\epsilon}\times\hat{\mathbf{k}})e^{-i\mathbf{k\cdot r}_j}|J\lambda\rangle.
\end{align}
and $\omega_\gamma$ is the energy of photon.

In the analysis of radiative transitions, we adopt the canonical
reference frame where the initial hadon is at rest, \(\mathbf{P_i} = 0\), and the final hadron carries momentum $\mathbf{P_f} = -\mathbf{k}$. The photon momentum is selected along the $z$-axis~($\mathbf{k} = k\hat{\mathbf{z}}$), and its  polarization vector is conventionally represented in a right-handed form, specifically as \(\epsilon = \frac{1}{\sqrt{2}}(1, i, 0)\).

To systematically evaluate the transition matrix elements,  we employ the method of multipole expansion for the plane wave \begin{align}
e^{-i\mathbf{k\cdot r}_j}=e^{-ikz_j}=\sum_l\sqrt{4\pi(2l+1)}(-i)^lj_l(kr_j)Y_{l0}(\Omega),
\end{align}
where \( j_l(x) \) denotes the spherical Bessel function of the first kind and \( Y_{lm}(\Omega) \) denotes the spherical harmonics. The matrix elements corresponding to the electric multipole transition amplitude and the magnetic transition amplitude are derived through angular momentum recoupling techniques~\cite{Deng:2016stx,Chen:2020jku,Li_2022}
\begin{align}
\mathcal{A}_\lambda^{\mathrm{E} l}  =&\sqrt{\frac{\omega_\gamma}{2}}\left\langle J^{\prime} \lambda^{\prime}\left|\sum_j(-i)^l B_l e_j j_{l+1}\left(k r_j\right) r_j Y_{l 1}\right| J \lambda\right\rangle \notag\\
& +\sqrt{\frac{\omega_\gamma}{2}}\left\langle J^{\prime} \lambda^{\prime}\left|\sum_j(-i)^l B_l e_j j_{l-1}\left(k r_j\right) r_j Y_{l 1}\right| J \lambda\right\rangle,\\
 \mathcal{A}_\lambda^{\mathrm{M} l}=&\sqrt{\frac{\omega_\gamma}{2}}\left\langle J^{\prime} \lambda^{\prime}\left|\sum_j(-i)^l C_l \frac{e_j}{2 m_j} j_{l-1}\left(k r_j\right) \sigma_j^{+} Y_{l-10} \right| J \lambda\right\rangle,
\end{align}
with the coefficients $B_l=\sqrt{2\pi l(l+1)/(2l+1)}$, $C_l=i\sqrt{8\pi(2l-1)}$, where $\sigma^+=(\sigma_x+i\sigma_y)/2$ represents the spin raising operator in the spherical tensor basis.

The electric and magnetic transition widths are given by
\begin{align}
\Gamma^{E l} & =\frac{|\boldsymbol{k}|^2}{\pi} \frac{2}{2 J_i+1} \frac{M_f}{M_i} \sum_{J_{zf}, J_{z i}}\left|\mathcal{A}_{J_{zf}, J_{z i}}^{E l}\right|^2, \\
\Gamma^{M l} & =\frac{|\boldsymbol{k}|^2}{\pi} \frac{2}{2 J_i+1} \frac{M_f}{M_i} \sum_{J_{zf}, J_{z i}}\left|\mathcal{A}_{J_{zf}, J_{z i}}^{M l}\right|^2,
\end{align}
where \( M_i \) and \( M_f \) are the masses of the initial and final charmonium states, respectively. Here, \( J_i \) denotes the total angular momentum of the initial hadron, while \( J_{zi} \) and \( J_{zf} \) represent the $z$-axis projections of the total angular momentum of the initial and final hadrons.

Contributions  from the higher angular momentum contributions are negligible compared with the lowest-order approximation~ ($l=1$). Therefore, the electric and magnetic transition widths reduce to
\begin{align}
\Gamma^{E1}= & \frac{4}{3} \alpha e_c^2 \omega_\gamma^3 \delta_{S S^{\prime}} \delta_{L L^{\prime} \pm 1} \max \left(L, L^{\prime}\right) \left(2 J^{\prime}+1\right) \notag\\
& \times\left\{\begin{array}{rrr}
L' & J' & S \\
J & L & 1
\end{array}\right\}^2\left\langle n^{\prime 2 S^{\prime}+1} L_{J^{\prime}}^{\prime}|r| n^{2 S+1} L_J\right\rangle^2,\label{E1formu}\\
\Gamma^{M1}= & \frac{4}{3}\frac{\alpha}{m_c^2} e_c^2 \omega_\gamma^3 \delta_{L L^{\prime}} \delta_{S S^{\prime} \pm 1} \frac{2 J^{\prime}+1}{2 L+1}\notag\\
&\times\left\langle n^{2 S^{\prime}+1} L_{J^{\prime}}^{\prime}\right| j_0\left(\omega_\gamma\frac{r}{2}\right)\left|n^{2 S+1} L_J\right\rangle^2,\label{M1formu}
\end{align}
where the charm quark charge $e_c=2/3$, the fine-structure constant $\alpha=1/137$, and $\{...\}$ represents the $6j$ symbol.

\section{Numerical results and discussion}\label{section3}

The present PDG table lists approximately 30 charmonium-like states~\cite{ParticleDataGroup:2022pth}. We also incorporate 17 bottomonium mass spectra as additional constraints to refine model parameters. Moreover, the ground states of charmed and charmed-strange quarkonium are included in the fitting procedure, since their wave functions are going to be used in the QPC model to obtain self-consistent predictions for the decay widths.
By fitting the mass spectrum, the model  parameters determined  are listed as follows:
\bqa
&m_{u/d}=0.346\mathrm{GeV},  m_s=0.548\mathrm{GeV}, m_c=1.748\mathrm{GeV}, m_b=5.077\mathrm{GeV},\nonumber\\
 &b=0.163\mathrm{GeV}^2, c=-0.409\mathrm{GeV},\sigma_0=2.02,s_0=2.017,\nonumber\\
 &\epsilon_{c}=-0.217,\epsilon_{t}=-0.716,\epsilon_{sov}=-0.212,\epsilon_{sos}=-0.031.
\eqa
This set of parameters are typically different from the original GI's values. With these set of parameters, the computed mass spectrum and candidate assignments are tabulated in Table~\ref{Pmassspectrum} and ~\ref{bbmassspectrum}.
When we calculate the open-charm strong decay widths in the QPC model, the production strength  from the vacuum is empirically  chosen to be $\gamma=6.3$ for $u\bar u$ or $d\bar d$ and $\gamma_s=6.3/\sqrt{3}$ for $s\bar s$~\cite{Chen:2018fsi}. The masses of the final states are chosen according to averages of the physical masses in the PDG table to ensure the correct threshold energies and phase space factors.
For the initial charmonium states, the average BW masses quoted in PDG table are used when the related candidates have been reported. For those states without candidates observed, we use theoretically predicted masses to compute the decay widths.
The total width is the sum over partial widths across all kinematically allowed decay channels, which   are also presented in Table~\ref{Pmassspectrum}.  In this calculation, the physical  charmed mesons $D_1(2420)$ and $D_1'(2430)$, as well as their charmed-strange counterparts $D_{s1}(2536)$ and $D_{s1}'(2460)$,  are interpreted as mixtures of the $1^1P_1$ and $1^3P_1$ configurations. The mixed states are defined as
\begin{align}
|P_1'\rangle&=\text{cos}~\theta~|1^1P_1\rangle+\text{sin}~\theta~|1^3P_1\rangle,\\
|P_1\rangle&=-\text{sin}~\theta~|1^1P_1\rangle+\text{cos}~\theta~|1^3P_1\rangle,
\end{align}
where the  mixing angle is empirically fixed at $\theta=-54.7^\circ$~\cite{Barnes:2002mu}.

While parameters of relativistic potential model are determined by fitting the mass spectrum, two significant uncertainties must be acknowledged: First, states with quantum numbers matching predictions of the quenched quark model do not necessarily correspond exclusively to conventional $q\bar q$ configurations. This ambiguity arises due to potential contamination from non-$q\bar q$ states  or near-threshold structures that may be interpreted as resonances in PDG listings. A second uncertainty stems from the extraction of mass parameters via Breit-Wigner~(BW) parametrization. The BW masses neither precisely correspond to poles mass from the scattering amplitudes nor bare masses from quark model, particularly when the narrow-resonance approximation breaks down for broad or overlapping states.

\begin{table}[t]
\centering
\renewcommand\arraystretch{1.}
\tabcolsep=0.3cm
\begin{tabular}{l|lcccccc}
\toprule[1.56pt]
Quarks & States & mass values & QPC widths &  PDG states & PDG masses& PDG widths & GI \\
\hline
\multirow{30}{*}{$c\bar{c}$}
& $1^1S_0$ &  2997  &   & $\eta_{c}$ &  2984 &  & 2974\\
& $2^1S_0$ &  3628 &   & $\eta_{c}'$ &  $3637$ & $20\pm 4$ &3630\\
& $3^1S_0$ &  4039 &  93 &  &   & &4068\\
& $4^1S_0$ &  4372 & 70  &  &   &   &4424\\
& $5^1S_0$ &  4663 & 117  &   &     &  &5018\\
\cline{2-8}
& $1^3S_1$ &  3122  &   & $J/\psi$ &  3096 &   &3092\\
& $2^3S_1$ &  3682 &   & $\psi(3686)$ &  $3686$ &  &3680\\
& $3^3S_1$ &  4076 & 65  & $\psi(4040)$ & 4040  & $84\pm 12$&4100\\
& $4^3S_1$ &  4401 & 86  & $\psi(4415)$ &  $4415\pm5$ & $110\pm22$ &4450\\
& $5^3S_1$ &  4687 & 71  & $\psi(4660)$ &  $4641\pm10$ & $73^{+13}_{-11}$  &4757\\
\cline{2-8}
& $1^3D_1$ &  3808  & 26   & $\psi(3770)$ &  $3773.7\pm 0.7$ & $27.2\pm 1.0$ &3818\\
& $2^3D_1$ &  4155 & 74  & $\psi(4160)$ &  $4191\pm 5$ & $69\pm 10$& 4194\\
& $3^3D_1$ &  4457 & 107  &   &    & & 4519\\
& $4^3D_1$ &  4728 & 120  &   &    &   &4812\\
& $5^3D_1$ &  4977 & 82  &   &     & &5080 \\
\cline{2-8}
& $1^3P_0$ &  3402  &   & $\chi_{c0}$ & 3414   &  &3443\\
& $2^3P_0$ &  3851 & 73  & $\chi_{c0}(3860)$? &  $3862^{+50}_{-35}$ & $200^{+180}_{-110}$& 3917\\
& $3^3P_0$ &  4204 & 85  &  &   &   &4291\\
& $4^3P_0$ &  4508 &  93 & $\chi_{c0}(4500)$ &  $4474\pm4$ & $77^{+12}_{-10}$ &4614\\
& $5^3P_0$ &  4781 & 79  & $\chi_{c0}(4700)$ &  $4694_{-5}^{+16}$  & $87^{+18}_{-10}$&4903\\
\cline{2-8}
& $1^3P_1$ &  3529 &   & $\chi_{c1}$ &  3510.67$\pm$0.05 &  &3508\\
& $2^3P_1$ &  3948 & 48  & $\chi_{c1}(4010)$ & $4012.5$  &  62.7 &3953\\
& $3^3P_1$ &  4286 & 19  & $\chi_{c1}(4274)$ &  $4286_{-9}^{+8}$ &$51\pm 7$& 4316\\
& $4^3P_1$ &  4581 & 242  & $\chi_{c1}(4685)$ &  $4684_{-17}^{+15}$ & $130\pm 40$&4633\\
& $5^3P_1$ &  4840 & 111  &   &    & &4920\\
\cline{2-8}
& $1^3P_2$ &  3567 &   & $\chi_{c2}$ &  3556.17$\pm$0.07  &  &3548\\
& $2^3P_2$ &  3971 & 20  & $\chi_{c2}(3930)$ & $3922.5\pm1.0$  & $35.2\pm 2.2$&3979\\
& $3^3P_2$ &  4304 & 46  & $X(4350)$? &  $4351\pm5$ & $13^{+18}_{-10}$ &4336\\
& $4^3P_2$ &  4595 & 91  &   &  & &4650 \\
& $5^3P_2$ &  4858 & 91  & &   & &4934\\
\cline{2-8}
& $1^1P_1$ &  3534 &   & $h_c$ &  3525.37$\pm$0.14 &   &3515\\
& $2^1P_1$ &  3946 & 63  & $X(3940)$? &  $3942\pm9$ & $37^{+27}_{-17}$& 3956\\
& $3^1P_1$ &  4283 & 31  &  &  &  &4318\\
& $4^1P_1$ &  4576 &  113 &   &   && 4634\\
& $5^1P_1$ &  4841 & 111  &  &  && 4920 \\
\toprule[1.6pt]
\end{tabular}
\caption{The mass spectrum and obtained strong decay widths of  charmonia  are compared with those of candidates in PDG. The unit is MeV. The candidates denoted with a question mark need further confirmations~\cite{ParticleDataGroup:2022pth}.}
\label{Pmassspectrum}
\end{table}

\begin{table}[t]
\centering
\renewcommand\arraystretch{1.}
\tabcolsep=0.3cm
\begin{tabular}{l|lcccc}
\toprule[1.56pt]
Quarks & States & mass values &  PDG states & PDG masses& GI \\
\hline
\multirow{30}{*}{$b\bar{b}$}
& $1^1S_0$ &  9449  &  $\eta_{b}(1S)$ &  $9398.7\pm2.0$     & 9431 \\
& $2^1S_0$ &  9998 &   $\eta_{b}(2S)$ &  $9999\pm 4$   & 9998 \\
& $3^1S_0$ &  10335 &    &   & 10354 \\
& $4^1S_0$ &  10601 &    &   &  10636  \\
& $5^1S_0$ &  10831 &     &     &  10881 \\
\cline{2-6}
& $1^3S_1$ &  9490  &   $\Upsilon(1S)$ &  $9460.4\pm0.10$ & 9472   \\
& $2^3S_1$ &  10014 &   $\Upsilon(2S)$  &  $10023.4\pm 0.5$ & 10015   \\
& $3^3S_1$ &  10345 &   $\Upsilon(3S)$ & $10355.1\pm0.5$  &  10365 \\
& $4^3S_1$ &  10609 &   $\Upsilon(4S)$ &  $10579.4\pm 1.2$ &10645   \\
& $5^3S_1$ &  10837 &   $\Upsilon(10860)$  & $10885.2^{+2.6}_{-1.6}$   &    10887  \\
\cline{2-6}
& $1^3D_1$ &  10134  &     &    &  10138  \\
& $2^3D_1$ &  10420 &      &     &  10442 \\
& $3^3D_1$ &  10662 &     &    &  10700 \\
& $4^3D_1$ &  10877 &      &    &  10930  \\
& $5^3D_1$ &  11074 &      &    &  11140   \\
\cline{2-6}
& $1^3P_0$ &  9856  &   $\chi_{b0}(1P)$  & $9859.44\pm0.73$   &  9850 \\
& $2^3P_0$ &  10218 &    $\chi_{b0}(2P)$ &  $10232.5\pm0.9$ &  10232 \\
& $3^3P_0$ &  10498 &      &   &  10529  \\
& $4^3P_0$ &  10736 &       &    &   10783 \\
& $5^3P_0$ &  10948 &       &     &  11001 \\
\cline{2-6}
& $1^3P_1$ &  9886 &    $\chi_{b1}(1P)$  &  $9892.78\pm0.57$ &  9876 \\
& $2^3P_1$ &  10239 &     $\chi_{b1}(2P)$  & $10255.46\pm0.72$  &  10249   \\
& $3^3P_1$ &  10514 &     $\chi_{b1}(3P)$  &  $10513.4\pm0.7$ &  10542 \\
& $4^3P_1$ &  10749 &     &    & 10792 \\
& $5^3P_1$ &  10959&       &    & 11017 \\
\cline{2-6}
& $1^3P_2$ &  9905 &     $\chi_{b1}(1P)$ &  $9912.21\pm0.57$  & 9896 \\
& $2^3P_2$ &  10252 &    $\chi_{b2}(2P)$  &  $10268.65\pm0.72$ &  10262 \\
& $3^3P_2$ &  10524 &     $\chi_{b2}(3P)$  &  $10524.0\pm0.8$ &  10551  \\
& $4^3P_2$ &  10757 &       &  &  10800 \\
& $5^3P_2$ &  10967 &    &   & 11024 \\
\cline{2-6}
& $1^1P_1$ &  9891 &    $h_b(1P)$ &  $9899.3\pm0.8$ &  9882  \\
& $2^1P_1$ &  10242 &    $h_b(2P)$  &  $10259.8\pm1.2$ &   10252 \\
& $3^1P_1$ &  10516 &     &  & 10544 \\
& $4^1P_1$ &  10751 &     &   &  10794\\
& $5^1P_1$ &  10961 &    &  & 11019 \\
\hline
\multirow{4}{*}{$c\bar{u}$}
& $1^1S_0$ &  1819  &  $D^0$ &  $1864.84\pm0.05$ &    1871  \\
& $1^3S_1$ &  2003 &   $D^{*0}(2007)$ &  $2006.85\pm 0.05$ & 2038 \\
& $1^3P_2$ &  2466 &  $D_2^*(2460)$  & $2461.1\pm0.8$  & 2501 \\
& $1^3P_1$ &  2438 &  $D_1(2420)$  &  $2422.1\pm0.6$ &   2490 \\
\hline
\multirow{4}{*}{$c\bar{s}$}
& $1^1S_0$ &  1947  &  $D_s^+$ &  $1968.35\pm0.07$ &   1975   \\
& $1^3S_1$ &  2119 &   $D_s^{*+}$ &  $2112.2\pm 0.4$ & 2124  \\
& $1^3P_2$ &  2579 &   $D_{s2}^{*+}(2573)$ &  $2569.1\pm0.8$ & 2591 \\
& $1^3P_1$ &  2544 &   $D_{s1}^{+}(2536)$ & $2535.11\pm0.06$  &  2570  \\
\toprule[1.6pt]
\end{tabular}
\caption{The mass spectrum of  bottomonia and some charmed/charmed-strange states  are compared with those of candidates in PDG. The unit is MeV~\cite{ParticleDataGroup:2022pth}.}
\label{bbmassspectrum}
\end{table}

 \subsection{$^3S_1$ and $^3D_1$ states}
A series of $J^{PC}=1^{--}$ charmonium states such as $\psi(3686)$, $\psi(3770)$, $\psi(4040)$, $\psi(4160)$, and $\psi(4415)$ have been observed and identified as the radial and orbital excitations $\psi(2S)$, $\psi(1D)$, $\psi(3S)$, $\psi(2D)$, and $\psi(4S)$, respectively, in different potential models~\cite{Godfrey:1985xj,Eichten:2007qx,Ebert:2002pp,Deng:2016stx}.
In recent years, high-precision data from $e^+e^-$ collider experiments at Belle, BaBar, BES, and CLEO-c have revealed additional resonance structures in the charmonium-like spectrum. The BES collaboration reported the hyperfine structure of the broad $Y(4260)$ state, identifying the lower $\psi(4230)$ and higher $\psi(4360)$ states. Very recently, BESIII  reported the observation of another $1^{--}$ state, $G(3900)$, in the $e^+e^-\rightarrow D\bar{D}$ process~\cite{BESIII:2024ths}.
Furthermore, the Belle have discovered another heavy $1^{--}$ state, $\psi(4660)$~\cite{Belle:2007umv},  which was confirmed by Babar~\cite{BaBar:2012hpr}. This $\psi(4660)$ state has been the subject of extensive theoretical debate with many non-$q\bar{q}$ interpretations including molecular states composed of $D_s^{(*)}\bar{D}_{s1}(2536)$, $D_s^{(*)}\bar{D}_{s2}(2573)$~\cite{He:2019csk,Wang:2020dya,Ke:2020eba}, $\psi(2S)f_0(980)$~\cite{Guo:2009id} and $\Lambda_c^+\bar{\Lambda}_c^-$~\cite{Cotugno:2009ys} or a $[cs][\bar{c}\bar{s}]$ tetraquark state~\cite{Zhang:2020gtx}.

In this calculation, we can  accommodate $J/\psi$, $\psi(3686)$, $\psi(4040)$, and $\psi(4415)$ as $\psi(nS)$ state with $n=1,\cdots, 4$, while  $\psi(3770)$ and   $\psi(4160)$ are consistent with the $\psi(1D)$ and $\psi(2D)$. This classification matches the assignments  in the GI model~\cite{Godfrey:1985xj}.
It is found that the calculated widths of $n^3S_1$ and $n^3D_1$ states are consistent with the extracted BW widths of the observed states. This implies the empirical choice of $\gamma=6.3$ is reasonable for most open-flavor decays.

The discrepancy emerges for the $\psi(5S)$ state.
The predicted mass of $\psi(5S)$ state is about 4757~MeV, which is about 110 MeV higher than the extracted BW mass of $\psi(4660)$. In the PDG review~\cite{ParticleDataGroup:2024cfk}, it is claimed that the $\psi(4660)$ challenges the quark model since its mass and decay properties are in conflicted with expectations. However, in this calculation, the mass of $\psi(5S)$ is $M_{\psi(5S)}=4687$~MeV, which is in agreement with the BW mass of $\psi(4660)$, which is $M_{\psi(4660)}=4641\pm 10$~MeV. This suggests that the $\psi(4660)$ could be identified as the $\psi(5S)$ state. Further evidence could be found from the strong decay widths.
The open-flavor strong widths obtained here is about 71~MeV, which is comparable with the PDG-averaged value $\Gamma_{\psi(4660)}=73^{+13}_{-11}$ MeV. Nevertheless, it is worth noting that the PDG-quoted widths are averaged over different experimental analyses, which deviate from each other. The recent observations by BESIII prefer a width larger than 100 MeV~\cite{BESIII:2023wqy,BESIII:2023cmv}, while the older measurements by Belle and Babar reported much narrower results~\cite{Belle:2014wyt,Belle:2008xmh}. Further experimental investigation are expected to determine the properties of this charmonium state more accurately.

In this classification of conventional charmonium, there is no room for the $G(3900)$, $\psi(4230)$ and $\psi(4360)$ states, strongly suggesting their non-$q\bar{q}$ nature. Notably, the $G(3900)$ has been proposed as a P-wave $D\bar{D}^*/\bar{D}D^*$ molecular state in ref.~\cite{Lin:2024qcq}. A recent $K$-matrix analysis of $e^+e^-\rightarrow D\bar{D}$ data also argued that  the $G(3900)$ does not originate from a conventional $c\bar{c}$ resonance but instead arises as a threshold enhancement linked to the $D\bar{D}^*/\bar{D}D^*$ threshold~\cite{Husken:2024hmi}. It is also worth mentioning that in the classic Cornell model~\cite{Eichten:1979ms}, a bump at about 3900~MeV in the $D\bar D$ invariant mass was predicted without introducing any new resonance other than the conventional charmonium states. Similarly, the  $\psi(4230)$ and $\psi(4360)$ states are usually understood in molecular interpretations due to the proximity of strongly coupled hadronic threshold near their observed masses~\cite{Liu:2013vfa}. These observations also implies that the traditional BW parametrization method might introduce more and more observed resonance shapes in the era of high-precision experimental investigation.
Other parametrization method beyond the BW form, such as coupled-channel analyses or effective range expansions, are essential for distinguishing true resonant states from threshold-induced enhancements. Such advances will clarify the fundamental nature of these exotic candidates.

 \subsection{$^3P_0$ states}

Long-standing challenges persist in the classification of $J^{PC}=0^{++}$ charmonium states.
The $\chi_{c0}(3915)$, first observed by Belle~\cite{Belle:2004lle} and later assigned   quantum numbers $0^{++}$ by Babar~\cite{BaBar:2012nxg}, initially appeared consistent with the $\chi_{c0}(2P)$ prediction of the  GI model. However, its absence in the $\gamma\gamma\rightarrow D\bar{D}$ process, where a narrow peak near 3930 MeV is identified as the $\chi_{c2}(2P)$ state, raised doubts about this assignment and inspired a suggestion of a broad structure at 3860 MeV linked to $\chi_{c0}(2P)$~\cite{Guo:2012tv}.
It was also pointed out that, if the helicity-2 dominance is abandoned, the peak at about 3930 MeV in $\gamma\gamma\rightarrow J/\psi\omega$  could be a $J^{PC}=2^{++}$ $\chi_{c2}(2P)$ state as in $\gamma\gamma\rightarrow D\bar{D}$ process~\cite{Zhou:2015uva}, which also implies the large helicity-0 component in this peak. Later, the Belle reported a broad $\chi_{c0}(3860)$ state in $e^+e^-\rightarrow J/\psi X$ process~\cite{Belle:2017egg}, which seems to be related to the board structure in the $\gamma\gamma\rightarrow D\bar{D}$ process. Different from the earlier analysis, LHCb made an  amplitude analysis of $B^+\rightarrow D^+D^-K^+$ by  modeling the data points about 3930 MeV in $D^+D^-$ invariant mass as contributed by nearly degenerate $\chi_{c0}(3915)$ and $\chi_{c2}(3930)$ in the BW parameterization form~\cite{LHCb:2020pxc}.  Their results do not accommodate the existence of a broad $\chi_{c0}(3860)$. These analyses add controversies at low mass scalar charmonia. In the higher energy region, the $\chi_{c0}(4500)$ and $\chi_{c0}(4700)$ states are observed by LHCb in $B^+\rightarrow J/\psi\phi K^+$~\cite{LHCb:2016axx}, both of which are more than 100 MeV lower than the predicted masses of conventional $c\bar{c}$ states, so they are also considered to be of exotic or non-$q\bar{q}$ origins in the PDG review.

In our calculation, the mass spectrum become denser than the predicted values in the original GI's work.
The predicted mass of $\chi_{c0}(2P)$ state is located at about 3851 MeV. This mass is similar to the results of other non-relativistic quark potential models~\cite{Barnes:2005pb,Li:2009zu}, but much lower than the original predicted mass of $3916$~MeV by GI's work.
The strong decay width of $\chi_{c0}(2P)$ obtained in the QPC model is quite large to about $\sim 73$ MeV,
so these results disfavors assigning
the $\chi_{c0}(3915)$ as the $\chi_{c0}(2P)$ state. To further investigate the possibility of  $\chi_{c0}(3915)$ as the $\chi_{c0}(2P)$ state,  we performed a constrained calculation using the same $\chi_{c0}(2P)$ wave function while fixing the mass at 3915 MeV by hand, and we found that  the open-charm decay width results in a much smaller value of $\sim 20$~MeV, which seems to be consistent with the measured width of $\chi_{c0}(3915)$. This comparison might imply that the calculated width is sensitive to nodal structures in the wave function, so the information of decay width is not sufficient for determining the property of $\chi_{c0}(3915)$.

For the excited states, the mass and width of the $\chi_{c0}(4P)$ state align closely with the observed $\chi_{c0}(4500)$ resonance reported by LHCb~\cite{LHCb:2016axx}. The calculated mass of the $\chi_{c0}(5P)$ state is 4781MeV, approximately 87~MeV above the BW mass of $\chi_{c0}(4700)$ quoted in the PDG table. However, it is important to note that in the $B^+\rightarrow J/\psi\phi K^+$ process, the upper limit of $J/\psi\phi$ energy is constrained by $m_{B^+}-m_{K^+}=4785.7$ MeV. This phase space restriction could distort the resonance line shape, potentially resulting in a lower extracted BW mass. It is also noticed that in an analysis of $B_s^0\to J/\psi\pi^+\pi^-K^+K^-$ by LHCb~\cite{LHCb:2020coc}, which has a larger phase space for $J/\psi\phi$, a similar structure, denoted as $X(4740)$, is observed and its BW mass and width are determined to be
$m_X=4741\pm 6\pm 6$~MeV and $\Gamma_X=53\pm 15\pm 11$~MeV. The measured values are more consistent with the results of $\chi_{c0}(5P)$ state obtained here, where $M_{\chi_{c0}(5P)}=4781$~MeV and $\Gamma_{\chi_{c0}(5P)}=79$~MeV.

At present, no confirmed candidate for the $\chi_{c0}(3P)$ charmonium state has been observed. In this work, the $\chi_{c0}(3P)$ is predicted to be located at approximately $4204$~MeV with a width of $85$~MeV. It is noticed that in the $B^+\rightarrow J/\psi\phi K^+$ process analyzed by LHCb~\cite{LHCb:2016axx}, where the higher-mass $\chi_{c0}(4500)$ and $\chi_{c0}(4700)$ states were identified, an unexplained structure (e.g., a kink or localized excess) appears near $4200 \text{MeV}$ in the $J/\psi\phi$ invariant mass spectrum. This feature is not accounted for in the model used in the experimental analysis in ref.~\cite{LHCb:2016axx}. The presence of this weak signal suggests the need for further experimental studies in this mass region to clarify its origin and explore potential resonance contributions.

 \subsection{$^3P_1$ states}

The quantum numbers of $^3P_1$ state are $J^{PC}=1^{++}$. The first excited $1^{++}$ charmonium-like state, the prominent  $\chi_{c1}(3872)$,  was the first heavy meson that could not be explained within the conventional $q\bar{q}$ quark model. The $\chi_{c1}(3872)$ lies near the $D^0\bar{D}^{*0}$ threshold and has an exceptionally narrow BW width of $\Gamma=1.19\pm 0.21$ MeV, strongly suggesting a non-$c\bar{c}$ origin. Additional higher $1^{++}$ charmonium-like states include the $\chi_{c1}(4140)$, $\chi_{c1}(4274)$ and $\chi_{c1}(4685)$. Recently, the LHCb collaboration reported the observation of the $\chi_{c1}(4010)$ in the $B^+\to D^{*+}D^-K^+$ process~\cite{LHCb:2024vfz}, further enriching the $1^{++}$ spectrum and highlighting its growing complexity.

In this analysis, the $\chi_{c1}(2P)$ state is tentatively associated with the recently observed $\chi_{c1}(4010)$ by LHCb~\cite{LHCb:2024vfz}. Although the predicted mass of $\chi_{c1}(2P)$ state is approximately 3948~MeV, which is about 60~MeV smaller than the BW mass of $\chi_{c1}(4010)$, a plausible explanation for this discrepancy is the inherent difference between the quenched model predictions and the BW mass determinations. Furthermore, interactions between the $\chi_{c1}(2P)$ and the $D\bar{D}$ scattering channels may play a role. In our prior work using the Lee-Friedrichs model~\cite{Zhou:2017dwj}, coupling the bare $\chi_{c1}(2P)$ state to $D\bar{D}$ channels was shown to generate two poles in the complex energy plane: one dynamically generated bound-state pole at the $D\bar{D}$ threshold and another shifted from the bare state into the complex energy plane. This framework aligns with recent findings in ref.~\cite{Deng:2023mza}, where the interplay with $\chi_{c1}(2P)$ produces a  $D\bar{D}$ lineshape characterized by a sharp threshold enhancement and a broad resonance near 3990~MeV. These features are attributed to the $\chi_{c1}(3872)$ and $\chi_{c1}(4010)$ respectively.

The $\chi_{c1}(3P)$ state in our calculation corresponds to the $\chi_{c1}(4274)$, while the $\chi_{c1}(4P)$ is predicted at $4581$~MeV with a broad width of $\sim242$~MeV within the QPC model. This $\chi_{c1}(4P)$ state might relate to the $\chi_{c1}(4685)$, but the mass difference of about 100 MeV is the largest one in the list. Since this $\chi_{c1}(4685)$ is only observed by one experiment, more experimental investigations are expected to verify or determine the physical parameters more accurately.
However, the $\chi_{c1}(4140)$ does not naturally align with this theoretical framework. Current experimental determinations of its properties show significant discrepancies: For instance, analyses of the $B^+\to J/\psi\phi K^+$ decay by the D0 and CMS collaborations report a narrow width of $\sim 20$~MeV~\cite{D0:2015nxw,CDF:2011pep}, whereas LHCb measures a substantially larger width of $\Gamma = 162 \pm 21^{+24}_{-49}$~MeV in the same channel~\cite{LHCb:2021uow}. These inconsistencies highlight the need for further experimental exploration to clarify the $\chi_{c1}(4140)$'s parameters and resolve its nature.

 \subsection{Radiative transitions}

We calculate the radiative E1 and M1 transitions of all relevant charmonium states. Results are presented in Tables \ref{E1-1}-\ref{E1-5}~(E1 transitions) and Tables \ref{M1-1}-\ref{M1-3}~(M1 transitions), with experimentally measured radiative partial widths included for comparison to theoretical predictions.

For the ground states, the computed E1 transitions of $\psi(1D)$, $\chi_{c0}(1P)$, $\chi_{c1}(1P)$, $\chi_{c2}(1P)$, and $h_{c}(1P)$ agree with the experimental values. Among excited states, only a few $S$-wave states have been measured.
For instance, we find  $\psi(2S)\to\gamma\chi_{cJ}$ yield $\Gamma(\psi(2S)\to\gamma\chi_{c0}/\chi_{c1}/\chi_{c2})=23/37/23$~keV respectively,
consistent with the experimental results in Table~\ref{E1-1}. For higher excitations,  only upper limits on transition rates exists for $\psi(3S)$ and  $\psi(4S)$.

For states lacking experimental data, we compare our results with those from established models~(e.g., LP and SP in ref.~\cite{Deng:2016stx}, NR and GI in ref.~\cite{Barnes:2005pb}) and SNR in ref.~\cite{Li:2009zu}.
Due to the ambiguity of $\chi_{c0}(2P)$, we provide two sets of predictions for  transitions involving the $\chi_{c0}(2P)$ state, corresponding to its tentative assignments as either the $\chi_{c0}(3860)$ or $\chi_{c0}(3915)$.

From the summary Tables \ref{E1-1}-\ref{E1-5} and Tables \ref{M1-1}-\ref{M1-3}, we identify several common principles governing dominant radiative transition channels:
\begin{enumerate}
  \item \textbf{ $S$-wave states:} The dominant E1 transitions of $n^1S_0$ and $n^3S_1$ occur between states with adjacent radial quantum number $n$. This pattern is systematically observed for $S$-wave charmonia~(Table \ref{E1-1}).
  \item \textbf{ $D$-wave states:} The dominant E1 transitions of $n^3D_1$ charmonium states follow $n^3D_1\to n^3P_0+\gamma$, where the final state retains the same radial quantum number and total spin $S$, but with the orbital angular momentum $L$ and total angular momentum $J$ reduced by one unit~(Table \ref{E1-2}).
  \item  \textbf{ $P$-wave states:} The dominant E1 transitions of $\chi_{cJ}(nP)$~($J=0,1,2$) usually proceed via  $\gamma\psi(nS)$, preserving the radial quantum number $n$~(Table \ref{E1-3} and \ref{E1-4}). { However,  exceptions arise  for the processes  $\chi_{c0}(4P)\to\gamma\psi(4S)$ and $\chi_{c0}(5P)\to\gamma\psi(5S)$, where the radiative partial widths are significantly suppressed. This suppression stems from the nodal structures of the initial- and final state wave functions.} At the same time, dominant channels of $h_c(nP)$ involve $\gamma\eta_c(nS)$ similarly conserving the radial quantum number $n$~(\ref{E1-5}).
  \item \textbf{M1 transitions:} The dominant channels predominantly feature transitions to ground-state mesons, as listed in \ref{M1-1}-\ref{M1-3}.
\end{enumerate}
These principles could be understood by viewing the eqs.~(\ref{E1formu}) and (\ref{M1formu}), which illustrates that the E1 transition mainly depends on the overlapping of the wave functions between the initial and final states, while the spherical Bessel function plays an important role in the M1 transition.

\section{Summary and conclusions}\label{summary}

We have reanalyze the mass spectrum of charmonia  in terms of the relativized potential model,  and computed  their open-charm strong decay widths and radiative transitions according to the wave functions obtained in the potential model. The mass spectrum of bottomonium state and some heavy-light ground states are also included in the fit to constrain the fit parameters. It is demonstrated that the new parameters could present a consistent description of the updated charmonium spectroscopy as well as their decay widths. We give a summary of the candidates of different sets of quantum numbers:

 (1)For the $J^{PC}=1^{--}$ state, the $J/\psi$, $\psi(3686)$, $\psi(4040)$, $\psi(4415)$ and $\psi(4660)$ could be associated with $c\bar{c}(n^3S_1)$ states of $n=1,\cdots,5$ respectively, while $\psi(3770)$ and $\psi(4160)$ are consistent with the $c\bar{c}(1^3D_1)$ and $c\bar{c}(2^3D_1)$.  Now the mass of  $c\bar{c}(5^3S_1)$ is consistent with the observed value of $\psi(4660)$.

 (2) For the $J^{PC}=0^{++}$ state, the $\chi_{c0}$, $\chi_{c0}(4500)$ and $\chi_{c0}(4700)$ are consistent with the $c\bar{c}(n^3P_0)$ states with $n=1,\ 4$ and $5$ respectively. The predicted mass of $c\bar{c}(2^3P_0)$ is located at about 3851~MeV, which is different from the observed $\chi_{c0}(3915)$. The candidate of $c\bar{c}(3^3P_0)$ is still absent. The kink structure of $J/\psi\phi$ invariant mass at about 4200~MeV is suggested to further explore and check whether there exist the $c\bar{c}(3^3P_0)$.

(3) For the $J^{PC}=1^{++}$ state, $\chi_{c1}(3510)$, $\chi_{c1}(4010)$, $\chi_{c1}(4274)$ and $\chi_{c1}(4685)$ are consistent with  the $c\bar{c}(n^3P_1)$ states with $n=1,\cdots, 4$  respectively. $\chi_{c1}(3872)$ could be regarded as the dynamically generated state due to the coupling of  $c\bar{c}(2^3P_1)$ to the $D\bar{D}$ channel, at the same time the mass of $c\bar{c}(2^3P_1)$ pole is also shifted to a higher mass due to coupled channel effect.

(4) For the $J^{PC}=2^{++}$ states, the $\chi_{c2}(3556)$ and $\chi_{c2}(3930)$ are associated with $c\bar{c}(1^3P_2)$ and $c\bar{c}(2^3P_2)$. At present, there is no other $2^{++}$ candidates reported in experiments.

In this calculation, most charmonium states could find their candidates observed in experiments.  The remaining puzzle is only the $c\bar{c}(2^3P_0)$, whose predicted mass and width  are  about 3851~MeV and 73~MeV respectively. This mass value is about 65~MeV lower than that of the GI's work~(about 3916~MeV), but it is close to those predicted value of non-relativistic potential model~\cite{Ebert:2002pp,Barnes:2005pb}.

Other states, such as $G(3900)$, $\chi_{c1}(4140)$,  $\psi(4230)$ and  $\psi(4360)$, are usually regarded as non-$q\bar{q}$ states so that could not be accommodated in this calculation.

Our analysis suggests that the GI model's original predictions for high charmonium states above the open-flavor threshold exhibit systematically larger mass splittings than experimental observations when using PDG-listed states as benchmarks. This discrepancy could be amended by recalibrating the model parameters to align theoretical predictions with data. Despite the challenges of describing the non-$q\bar{q}$ states, the quenched quark model remains to be a valuable tool for rapidly categorizing charmonium states and distinguishing conventional configurations in the experiments.

\begin{acknowledgments}
 This work is supported by China National Natural Science Foundation under contract No. 11975075 and No. 12375132.
\end{acknowledgments}

\begin{minipage}{\textwidth}
\begin{minipage}[t]{0.48\textwidth}
\vspace{0pt}
\makeatletter\def\@captype{table}
\begin{tabular}{c|ccc}
\toprule[1.5pt]
  Initial state & Final state & $\Gamma_i$(Our Result) & $\Gamma_{\text{exp}}$    \\
\hline
$\eta_c(2S)$ & $\gamma h_c(1P)$ & 33 \\
\hline
$\eta_c(3S)$ & $\gamma h_c(1P)$ & 43 \\
             & $\gamma h_c(2P)$ & 45 \\
\hline
$\eta_c(4S)$ & $\gamma h_c(1P)$ & 18 \\
             & $\gamma h_c(2P)$ & 43 \\
             & $\gamma h_c(3P)$ & 69 \\
\hline
$\eta_c(5S)$ & $\gamma h_c(1P)$ & 11 \\
             & $\gamma h_c(2P)$ & 17 \\
             & $\gamma h_c(3P)$ & 41 \\
             & $\gamma h_c(4P)$ & 88 \\
\hline
$\psi(2S)$   & $\gamma \chi_{c0}(1P)$ & 21 & 29$\pm$2 \\
             & $\gamma \chi_{c1}(1P)$ & 35 & 29$\pm$2 \\
             & $\gamma \chi_{c2}(1P)$ & 22 & 27$\pm$2 \\
\hline
$\psi(3S)$   & $\gamma \chi_{c0}(1P)$ &  4 \\
             & $\gamma \chi_{c0}(2P)$ & 15/4 \\
             & $\gamma \chi_{c1}(1P)$ & 1  & \multirow{2}{*}{$<$326} \\
             & $\gamma \chi_{c1}(2P)$ & 3$\times 10^{-1}$ \\
             & $\gamma \chi_{c2}(1P)$ & 14 & \multirow{2}{*}{$<$480} \\
             & $\gamma \chi_{c2}(2P)$ & 51 \\
\hline
$\psi(4S)$   & $\gamma \chi_{c0}(1P)$ & 3 \\
             & $\gamma \chi_{c0}(2P)$ & 6/4 \\
             & $\gamma \chi_{c0}(3P)$ & 45 \\
             & $\gamma \chi_{c1}(1P)$ & 1 & \multirow{3}{*}{$<106$}\\
             & $\gamma \chi_{c1}(2P)$ & 1 \\
             & $\gamma \chi_{c1}(3P)$ & 60 \\
             & $\gamma \chi_{c2}(1P)$ & 9 & \multirow{3}{*}{$<528$}\\
             & $\gamma \chi_{c2}(2P)$ & 31 \\
             & $\gamma \chi_{c2}(3P)$ & 72 \\
\hline
$\psi(5S)$   & $\gamma \chi_{c0}(1P)$ & 2 \\
             & $\gamma \chi_{c0}(2P)$ & 3/2 \\
             & $\gamma \chi_{c0}(3P)$ & 4 \\
             & $\gamma \chi_{c0}(4P)$ & 32 \\
             & $\gamma \chi_{c1}(1P)$ & 1 \\
             & $\gamma \chi_{c1}(2P)$ & 5$\times 10^{-1}$ \\
             & $\gamma \chi_{c1}(3P)$ & 2 \\
             & $\gamma \chi_{c2}(1P)$ & 5 \\
             & $\gamma \chi_{c2}(2P)$ & 9 \\
             & $\gamma \chi_{c2}(3P)$ & 14 \\
             & $\gamma \chi_{c2}(4P)$ & 7 \\
\toprule[1.5pt]
\end{tabular}
\caption{The E1 radiative transitions of $S$-wave states.
The unit is keV.}\label{E1-1}
\end{minipage}
\begin{minipage}[t]{0.48\textwidth}
\vspace{0pt}
\makeatletter\def\@captype{table}
\begin{tabular}{c|ccc}
\toprule[1.5pt]
  Initial state & Final state & $\Gamma_i$(Our Result) & $\Gamma_{\text{exp}}$   \\
\hline
$\psi(1D)$   & $\gamma \chi_{c0}(1P)$ & 229 & 188$\pm$23 \\
             & $\gamma \chi_{c1}(1P)$ & 102 & 68$\pm$9 \\
             & $\gamma \chi_{c2}(1P)$ & 4   & $<$18 \\
\hline
$\psi(2D)$   & $\gamma \chi_{c0}(1P)$ & 117 \\
             & $\gamma \chi_{c0}(2P)$ & 288/154 \\
             & $\gamma \chi_{c1}(1P)$ & 1 \\
             & $\gamma \chi_{c1}(2P)$ & 59 \\
             & $\gamma \chi_{c2}(1P)$ & 2$\times 10^{-1}$ \\
             & $\gamma \chi_{c2}(2P)$ & 16 \\
\hline
$\psi(3D)$   & $\gamma \chi_{c0}(1P)$ & 62 \\
             & $\gamma \chi_{c0}(2P)$ & 65/46 \\
             & $\gamma \chi_{c0}(3P)$ & 192 \\
             & $\gamma \chi_{c1}(1P)$ & 1$\times 10^{-1}$ \\
             & $\gamma \chi_{c1}(2P)$ & 2$\times 10^{-1}$ \\
             & $\gamma \chi_{c1}(3P)$ & 81 \\
             & $\gamma \chi_{c2}(1P)$ & 3$\times 10^{-1}$ \\
             & $\gamma \chi_{c2}(2P)$ & 3$\times 10^{-1}$ \\
             & $\gamma \chi_{c2}(3P)$ & 4 \\
\hline
$\psi(4D)$   & $\gamma \chi_{c0}(1P)$ & 46 \\
             & $\gamma \chi_{c0}(2P)$ & 42/33 \\
             & $\gamma \chi_{c0}(3P)$ & 50 \\
             & $\gamma \chi_{c0}(4P)$ & 273 \\
             & $\gamma \chi_{c1}(1P)$ & 1 \\
             & $\gamma \chi_{c1}(2P)$ & 2$\times 10^{-1}$ \\
             & $\gamma \chi_{c1}(3P)$ & 2$\times 10^{-4}$ \\
             & $\gamma \chi_{c1}(4P)$ & 2 \\
             & $\gamma \chi_{c2}(1P)$ & 4$\times 10^{-1}$ \\
             & $\gamma \chi_{c2}(2P)$ & 4$\times 10^{-1}$ \\
             & $\gamma \chi_{c2}(3P)$ & 3$\times 10^{-1}$ \\
             & $\gamma \chi_{c2}(4P)$ & 4 \\
\hline
$\psi(5D)$   & $\gamma \chi_{c0}(1P)$ & 37 \\
             & $\gamma \chi_{c0}(2P)$ & 31/26 \\
             & $\gamma \chi_{c0}(3P)$ & 33 \\
             & $\gamma \chi_{c0}(4P)$ & 48 \\
             & $\gamma \chi_{c0}(5P)$ & 507 \\
             & $\gamma \chi_{c1}(1P)$ & 1 \\
             & $\gamma \chi_{c1}(2P)$ & 1 \\
             & $\gamma \chi_{c1}(3P)$ & 3$\times 10^{-1}$ \\
             & $\gamma \chi_{c1}(4P)$ & 8$\times 10^{-2}$ \\
             & $\gamma \chi_{c1}(5P)$ & 65 \\
             & $\gamma \chi_{c2}(1P)$ & 3$\times 10^{-1}$ \\
             & $\gamma \chi_{c2}(2P)$ & 4$\times 10^{-1}$ \\
             & $\gamma \chi_{c2}(3P)$ & 3$\times 10^{-1}$ \\
             & $\gamma \chi_{c2}(4P)$ & 3$\times 10^{-1}$ \\
             & $\gamma \chi_{c2}(5P)$ & 4 \\
\toprule[1.5pt]
\end{tabular}
\caption{The E1 radiative transitions of $^3D_1$ charmonium states. The unit is keV.}\label{E1-2}
\end{minipage}
\end{minipage}

\begin{table}[H]
\centering
\renewcommand\arraystretch{1.05}
\tabcolsep=0.3cm
\begin{tabular}{c|cccccc}
\toprule[1.5pt]
  Initial state & Final state & $\Gamma_i$(Our Result) & $\Gamma_i$(LP/SP)\cite{PhysRevD.95.034026} & $\Gamma_i$(NR/GI)\cite{PhysRevD.72.054026} & $\Gamma_i$(SNR$_{0/1}$)\cite{Li:2009zu}  & $\Gamma_{\text{exp}}$   \\
\hline
 $\chi_{c0}(1P)$ & $\gamma J/\psi(1S)$ & 145 &  &  &  & 151$\pm$18 \\
\hline
 $\chi_{c0}(2P)$ & $\gamma J/\psi(1S)$ &  43/56 & 4.0/1.5 & 56/1.3  & 74/9.3\\
                 & $\gamma \psi(2S)$ &  80/197 & 108/89 & 64/135 & 61/44      \\
                 & $\gamma \psi_1(1D)$ &  15/71 & 21/12 & 13/51 &...     \\
 \hline
 $\chi_{c0}(3P)$ & $\gamma J/\psi(1S)$ & 51  & 0.14/0.08 & 27/1.5& ...  \\
                       & $\gamma \psi(2S)$ & 27  & 13/6.9  & 32/0.045 & ... \\
                       & $\gamma \psi(3S)$ & 125  & 214/56  & 109/145 & ...  \\
                       & $\gamma \psi_1(1D)$ & 40  & 0.31/0.44  & 0.037/9.7  & ... \\
                       & $\gamma \psi_1(2D)$ & 1$\times 10^{-1}$ & ...  & ...  & ...   \\
 \hline
 $\chi_{c0}(4P)$     & $\gamma J/\psi(1S)$ & 46  \\
                     & $\gamma \psi(2S)$ & 23  \\
                     & $\gamma \psi(3S)$ & 29  \\
                     & $\gamma \psi(4S)$ & 9  \\
                     & $\gamma \psi_1(1D)$ & 10  \\
                     & $\gamma \psi_1(2D)$ & 23  \\
                     & $\gamma \psi_1(3D)$ & 4$\times 10^{-1}$  \\
 \hline
 $\chi_{c0}(5P)$       & $\gamma J/\psi(1S)$ & 39  \\
                       & $\gamma \psi(2S)$ & 19  \\
                       & $\gamma \psi(3S)$ & 17  \\
                       & $\gamma \psi(4S)$ & 11  \\
                       & $\gamma \psi(5S)$ & 9  \\
                       & $\gamma \psi_1(1D)$ & 4  \\
                       & $\gamma \psi_1(2D)$ & 7  \\
                       & $\gamma \psi_1(3D)$ & 20  \\
 \hline
 $\chi_{c1}(1P)$ & $\gamma J/\psi(1S)$ & 339 &  &  &  & 288$\pm$24  \\
 \hline
 $\chi_{c1}(2P)$ & $\gamma J/\psi(1S)$ & 54 & 64/68 & 71/14 & 83/45   \\
                & $\gamma \psi(2S)$ & 416 & 129/145 & 183/183 & 103/60 \\
                & $\gamma \psi_1(1D)$ & 51 & 8.6/10.8 & 22/21 & ... \\
                & $\gamma \psi_2(1D)$ & 75 & 2.8/3.4 & 35/18 & ...  \\
 \hline
 $\chi_{c1}(3P)$ & $\gamma J/\psi(1S)$ & 15 & 36/33 & 31/2.2 & ... \\
                 & $\gamma \psi(2S)$ & 21 & 78/63 & 45/8.9 & ... \\
                 & $\gamma \psi(3S)$ & 107 & 305/111 & 303/181 & ... \\
                 & $\gamma \psi_1(1D)$ & 2$\times 10^{-2}$  & 4.4/2.7 & 0.0014/0.39 & ... \\
                 & $\gamma \psi_1(2D)$ & 7 & 8.6/0 & 19/15 & ... \\
                 & $\gamma \psi_2(1D)$ & 4 & 0.13/0.09 & 0.0035/4.6 & ... \\
                 & $\gamma \psi_2(2D)$ & 33 & 22/11 & 58/35 & ... \\
 \hline
 $\chi_{c1}(4P)$ & $\gamma J/\psi(1S)$ & 10  \\
                 & $\gamma \psi(2S)$ & 14  \\
                 & $\gamma \psi(3S)$ & 31  \\
                 & $\gamma \psi(4S)$ & 602  \\
                 & $\gamma \psi_1(1D)$ & 9$\times 10^{-4}$   \\
                 & $\gamma \psi_1(2D)$ & 9$\times 10^{-4}$   \\
                 & $\gamma \psi_1(3D)$ & 154  \\
                 & $\gamma \psi_2(1D)$ & 2  \\
                 & $\gamma \psi_2(2D)$ & 12  \\
                 & $\gamma \psi_2(3D)$ & 405  \\
\hline
 $\chi_{c1}(5P)$ & $\gamma J/\psi(1S)$ & 5  \\
                 & $\gamma \psi(2S)$ & 6  \\
                 & $\gamma \psi(3S)$ & 8  \\
                 & $\gamma \psi(4S)$ & 9  \\
                 & $\gamma \psi(5S)$ & 361  \\
                 & $\gamma \psi_1(1D)$ & 5$\times 10^{-3}$   \\
                 & $\gamma \psi_1(2D)$ & 3$\times 10^{-2}$   \\
                 & $\gamma \psi_1(3D)$ & 1$\times 10^{-2}$   \\
                 & $\gamma \psi_1(4D)$ & 30  \\
                 & $\gamma \psi_2(1D)$ & 1  \\
                 & $\gamma \psi_2(2D)$ & 2  \\
                 & $\gamma \psi_2(3D)$ & 7  \\
                 & $\gamma \psi_2(4D)$ & 55 \\
\toprule[1.5pt]
\end{tabular}
\caption{The E1 radiative transitions  of $^3P_0$ and $^3P_1$ charmonium states. The unit is keV.}\label{E1-3}
\end{table}

\begin{table}[H]
\centering
\renewcommand\arraystretch{1.15}
\tabcolsep=0.3cm
\begin{tabular}{c|cccccc}
\toprule[1.5pt]
  Initial state & Final state & $\Gamma_i$(Our Result)& $\Gamma_i$(LP/SP)\cite{PhysRevD.95.034026} & $\Gamma_i$(NR/GI)\cite{PhysRevD.72.054026} & $\Gamma_i$(SNR$_{0/1}$)\cite{Li:2009zu} & $\Gamma_{\text{exp}}$   \\
\hline
$\chi_{c2}(1P)$ & $\gamma J/\psi(1S)$ & 493  &  &  &  & 386$\pm$33  \\
\hline
$\chi_{c2}(2P)$ & $\gamma J/\psi(1S)$ & 103  & 118/119 & 81/53 & 101/109 \\
                & $\gamma \psi(2S)$ & 131 & 146/163 & 304/207 & 225/100 \\
                & $\gamma \psi_1(1D)$ & 4$\times 10^{-1}$  & 0.47/0.62 & 1.9/1.0 & ... \\
                & $\gamma \psi_2(1D)$ & 1 & 3.3/4.1  & 17/5.6 & ... \\
                & $\gamma \psi_3(1D)$ & 6 & 20/24 & 88/29  & ... \\
\hline
  $\chi_{c2}(3P)$ & $\gamma J/\psi(1S)$ & 70 & 83/69 & 34/19 & ... \\
                  & $\gamma \psi(2S)$ & 52  & 116/90 & 55/30 & ... \\
                  & $\gamma \psi(3S)$ & 325 & 306/121 & 509/199 & ... \\
                  & $\gamma \psi_1(1D)$ & 1$\times 10^{-2}$  & 1.9/1.0  & $\sim$0/0.001 & ... \\
                  & $\gamma \psi_1(2D)$ & 4$\times 10^{-1}$  & 0.55/0.004 & 2.1/0.77 & ... \\
                  & $\gamma \psi_2(1D)$ & 9$\times 10^{-2}$  & 4.6/2.5 & 0.0091/0.13 & ... \\
                  & $\gamma \psi_2(2D)$ & 9 & 18/10  & 31/9.9 & ... \\
                  & $\gamma \psi_3(1D)$ & 4 &  15/10 & 0.049/6.8 & ... \\
                  & $\gamma \psi_3(2D)$ & 47 & 116/64  & 148/51 & ... \\
\hline
  $\chi_{c2}(4P)$ & $\gamma J/\psi(1S)$ & 48  \\
                  & $\gamma \psi(2S)$ & 30  \\
                  & $\gamma \psi(3S)$ & 42  \\
                  & $\gamma \psi(4S)$ & 151   \\
                  & $\gamma \psi_1(1D)$ & 8$\times 10^{-3}$   \\
                  & $\gamma \psi_1(2D)$ & 2$\times 10^{-2}$   \\
                  & $\gamma \psi_1(3D)$ & 1   \\
                  & $\gamma \psi_2(1D)$ & 3$\times 10^{-2}$   \\
                  & $\gamma \psi_2(2D)$ & 2$\times 10^{-1}$   \\
                  & $\gamma \psi_2(3D)$ & 12  \\
                  & $\gamma \psi_3(1D)$ & 1  \\
                  & $\gamma \psi_3(2D)$ & 6  \\
                  & $\gamma \psi_3(3D)$ & 60  \\
\hline
 $\chi_{c2}(5P)$  & $\gamma J/\psi(1S)$ & 37   \\
                  & $\gamma \psi(2S)$ & 21   \\
                  & $\gamma \psi(3S)$ & 24   \\
                  & $\gamma \psi(4S)$ & 22   \\
                  & $\gamma \psi(5S)$ & 369   \\
                  & $\gamma \psi_1(1D)$ & 5$\times 10^{-3}$   \\
                  & $\gamma \psi_1(2D)$ & 1$\times 10^{-2}$  \\
                  & $\gamma \psi_1(3D)$ & 3$\times 10^{-2}$    \\
                  & $\gamma \psi_1(4D)$ & 1  \\
                  & $\gamma \psi_2(1D)$ & 2$\times 10^{-2}$    \\
                  & $\gamma \psi_2(2D)$ & 4$\times 10^{-4}$    \\
                  & $\gamma \psi_2(3D)$ & 2$\times 10^{-1}$   \\
                  & $\gamma \psi_2(4D)$ & 14   \\
                  & $\gamma \psi_3(1D)$ & 1   \\
                  & $\gamma \psi_3(2D)$ & 2   \\
                  & $\gamma \psi_3(3D)$ & 7   \\
                  & $\gamma \psi_3(4D)$ & 69   \\
\toprule[1.5pt]
\end{tabular}
\caption{The E1 radiative transitions of $^3P_2$ charmonium states. The unit is keV.}\label{E1-4}
\end{table}

\begin{table}[H]
\centering
\renewcommand\arraystretch{1.15}
\tabcolsep=0.3cm
\begin{tabular}{c|cccccc}
\toprule[1.5pt]
  Initial state & Final state & $\Gamma_i$(Our Result) & $\Gamma_i$(LP/SP)\cite{PhysRevD.95.034026} & $\Gamma_i$(NR/GI)\cite{PhysRevD.72.054026} & $\Gamma_i$(SNR$_{0/1}$)\cite{Li:2009zu} & $\Gamma_{\text{exp}}$    \\
\hline
 $h_c(1P)$       & $\gamma \eta_c(1S)$ & 559 &   &   &   & $468_{-188}^{+210}$  \\
\hline
 $h_c(2P)$       & $\gamma \eta_c(1S)$ & 204 & 135/134  & 140/85  & 134/250 \\
                 & $\gamma \eta_c(2S)$ & 251 & 160/176  & 280/218  & 309/108  \\
                 & $\gamma \eta_{c2}(1D)$ & 24 & 25/25  & 60/27  & ... \\
\hline
 $h_c(3P)$       & $\gamma \eta_c(1S)$ &  124  & 90/77 & 72/38 \\
                 & $\gamma \eta_c(2S)$ & 68  & 124/96 & 75/43  \\
                 & $\gamma \eta_c(3S)$ & 224  & 237/146 & 276/208 \\
                 & $\gamma \eta_{c2}(1D)$ & 6  &  15/8.7 & 0.16/5.7\\
                 & $\gamma \eta_{c2}(2D)$ & 43  & 93/47 & 99/48 \\
 \hline
  $h_c(4P)$       & $\gamma \eta_c(1S)$ & 91 \\
                  & $\gamma \eta_c(2S)$ & 41 \\
                  & $\gamma \eta_c(3S)$ & 43 \\
                  & $\gamma \eta_c(4S)$ & 203 \\
                  & $\gamma \eta_{c2}(1D)$ & 2 \\
                  & $\gamma \eta_{c2}(2D)$ & 10 \\
                  & $\gamma \eta_{c2}(3D)$ & 57 \\
\hline
 $h_c(5P)$        & $\gamma \eta_c(1S)$ & 72  \\
                  & $\gamma \eta_c(2S)$ & 29  \\
                  & $\gamma \eta_c(3S)$ & 26  \\
                  & $\gamma \eta_c(4S)$ & 29  \\
                  & $\gamma \eta_c(5S)$ & 190  \\
                  & $\gamma \eta_{c2}(1D)$ & 1  \\
                  & $\gamma \eta_{c2}(2D)$ & 3  \\
                  & $\gamma \eta_{c2}(3D)$ & 12  \\
                  & $\gamma \eta_{c2}(4D)$ & 68  \\
\toprule[1.5pt]
\end{tabular}
\caption{The E1 radiative transitions of $^1P_1$ charmonium states. The unit is keV.}\label{E1-5}
\end{table}

\begin{minipage}{\textwidth}
\begin{minipage}[t]{0.48\textwidth}
\vspace{0pt}
\makeatletter\def\@captype{table}
\begin{tabular}{c|ccc}
\toprule[1.2pt]
  Initial state & Final state & $\Gamma_i$(Our Result) & $\Gamma_{\text{exp}}$   \\
\hline
$\eta_c(2S)$ & $\gamma J/\psi(1S)$ & 9 & $<$188 \\
\hline
$\eta_c(3S)$ & $\gamma J/\psi(1S)$ & 12 \\
             &   $\gamma \psi(2S)$ & 1 \\
\hline
$\eta_c(4S)$ & $\gamma J/\psi(1S)$ & 14 \\
             &   $\gamma \psi(2S)$ & 3 \\
             &   $\gamma \psi(3S)$ & 2$\times 10^{-1}$  \\
\hline
$\eta_c(5S)$ & $\gamma J/\psi(1S)$ & 13 \\
             &   $\gamma \psi(2S)$ & 4 \\
             &   $\gamma \psi(3S)$ & 1 \\
             &   $\gamma \psi(4S)$ & 1$\times 10^{-1}$  \\
             &   $\gamma \psi(5S)$ & 5$\times 10^{-2}$  \\
\hline
$\psi(1S)$   &  $\gamma \eta_c(1S)$ & 2 \\
\hline
$\psi(2S)$   &  $\gamma \eta_c(1S)$ & 23  & 1 \\
             &  $\gamma \eta_c(2S)$ & 2$\times 10^{-1}$  & (2$\pm$1)$\times 10^{-1}$  \\
\hline
$\psi(3S)$   &  $\gamma \eta_c(1S)$ & 29 \\
             &  $\gamma \eta_c(2S)$ & 3 \\
             &  $\gamma \eta_c(3S)$ & 1$\times 10^{-7}$  \\
\hline
$\psi(4S)$   &  $\gamma \eta_c(1S)$ & 51 \\
             &  $\gamma \eta_c(2S)$ & 7 \\
             &  $\gamma \eta_c(3S)$ & 3 \\
             &  $\gamma \eta_c(4S)$ & 1$\times 10^{-2}$  \\
\hline
$\psi(5S)$   &  $\gamma \eta_c(1S)$ & 54 \\
             &  $\gamma \eta_c(2S)$ & 8 \\
             &  $\gamma \eta_c(3S)$ & 3 \\
             &  $\gamma \eta_c(4S)$ & 1 \\
\hline
$\psi(2D)$   &  $\gamma \eta_{c2}(1D)$ & 8$\times 10^{-2}$  \\
             &  $\gamma \eta_{c2}(2D)$ & 2$\times 10^{-3}$  \\
\hline
$\psi(3D)$   &  $\gamma \eta_{c2}(1D)$ & 3$\times 10^{-1}$  \\
             &  $\gamma \eta_{c2}(2D)$ & 8$\times 10^{-2}$  \\
\hline
$\psi(4D)$   &  $\gamma \eta_{c2}(1D)$ & 5$\times 10^{-1}$  \\
             &  $\gamma \eta_{c2}(2D)$ & 3$\times 10^{-1}$  \\
             &  $\gamma \eta_{c2}(3D)$ & 6$\times 10^{-2}$  \\
\hline
$\psi(5D)$   &  $\gamma \eta_{c2}(1D)$ & 1 \\
             &  $\gamma \eta_{c2}(2D)$ & 1 \\
             &  $\gamma \eta_{c2}(3D)$ & 3$\times 10^{-1}$  \\
             &  $\gamma \eta_{c2}(4D)$ & 4$\times 10^{-2}$  \\
\toprule[1.5pt]
\end{tabular}
\caption{The M1 radiative transitions of $^3S_1$ and $^3D_1$ charmonium states. The unit is keV.}\label{M1-1}
\end{minipage}
\begin{minipage}[t]{0.48\textwidth}
\vspace{0pt}
\makeatletter\def\@captype{table}
\begin{tabular}{c|ccc}
\toprule[1.2pt]
  Initial state & Final state & $\Gamma_i$(Our Result)& $\Gamma_i$(NR/GI)\cite{PhysRevD.72.054026}     \\
\hline
 $\chi_{c0}(2P)$  & $\gamma h_c(1P)$ & 3/5 &  0.19/0.50   \\
 \hline
 $\chi_{c0}(3P)$  & $\gamma h_c(1P)$ &  6   \\
                  & $\gamma h_c(2P)$ &  2   \\
 \hline
 $\chi_{c0}(4P)$  & $\gamma h_c(1P)$ &  7  \\
                  & $\gamma h_c(2P)$ &  3  \\
                  & $\gamma h_c(3P)$ &  1  \\
\hline
 $\chi_{c0}(5P)$  & $\gamma h_c(1P)$ &  7  \\
                  & $\gamma h_c(2P)$ &  4  \\
                  & $\gamma h_c(3P)$ &  2  \\
                  & $\gamma h_c(4P)$ &  2$\times 10^{-1}$   \\
\hline
 $\chi_{c1}(2P)$  & $\gamma h_c(1P)$ & 1  & 0.51/0.045   \\
                  & $\gamma h_c(2P)$ & 5$\times 10^{-1}$   \\
\hline
 $\chi_{c1}(3P)$  & $\gamma h_c(1P)$ & 2$\times 10^{-1}$   \\
                  & $\gamma h_c(2P)$ & 2$\times 10^{-1}$   \\
                  & $\gamma h_c(3P)$ & 4$\times 10^{-5}$   \\
\hline
 $\chi_{c1}(4P)$  & $\gamma h_c(1P)$ & 1  \\
                  & $\gamma h_c(2P)$ & 1  \\
                  & $\gamma h_c(3P)$ & 2  \\
                  & $\gamma h_c(4P)$ & 2  \\
\hline
 $\chi_{c1}(5P)$  & $\gamma h_c(1P)$ & 5$\times 10^{-1}$   \\
                  & $\gamma h_c(2P)$ & 4$\times 10^{-1}$   \\
                  & $\gamma h_c(3P)$ & 4$\times 10^{-1}$   \\
                  & $\gamma h_c(4P)$ & 3$\times 10^{-1}$   \\
                  & $\gamma h_c(5P)$ & 3$\times 10^{-4}$   \\
\hline
$\chi_{c2}(1P)$   & $\gamma h_c(1P)$ & 1$\times 10^{-1}$  &    \\
\hline
$\chi_{c2}(2P)$   & $\gamma h_c(1P)$ & 1 & 0.90/1.3   \\
\hline
$\chi_{c2}(3P)$   & $\gamma h_c(1P)$ & 3  \\
                  & $\gamma h_c(2P)$ & 1  \\
                  & $\gamma h_c(3P)$ & 2$\times 10^{-2}$   \\
\hline
  $\chi_{c2}(4P)$ & $\gamma h_c(1P)$ & 4 \\
                  & $\gamma h_c(2P)$ & 2 \\
                  & $\gamma h_c(3P)$ & 1 \\
                  & $\gamma h_c(4P)$ & 1$\times 10^{-2}$  \\
\hline
 $\chi_{c2}(5P)$  & $\gamma h_c(1P)$ & 4  \\
                  & $\gamma h_c(2P)$ & 3  \\
                  & $\gamma h_c(3P)$ & 2  \\
                  & $\gamma h_c(4P)$ & 1  \\
                  & $\gamma h_c(5P)$ & 8$\times 10^{-3}$   \\
\toprule[1.5pt]
\end{tabular}
\caption{The M1 radiative transitions of $^3P_0$, $^3P_1$ and $^3P_2$ charmonium states. The unit is keV.}\label{M1-2}
\end{minipage}
\end{minipage}

\begin{table}[H]
\centering
\renewcommand\arraystretch{1.15}
\tabcolsep=0.3cm
\begin{tabular}{c|ccc}
\toprule[1.2pt]
  Initial state & Final state & $\Gamma_i$(Our Result) & $\Gamma_i$(NR/GI)\cite{PhysRevD.72.054026}     \\
\hline
 $h_c(1P)$        & $\gamma \chi_{c0}(1P)$ & 1 &  \\
                  & $\gamma \chi_{c1}(1P)$ & 5$\times 10^{-3}$  &  \\
\hline
 $h_c(2P)$        & $\gamma \chi_{c0}(1P)$ & 6 & 0.21/1.5 \\
                  & $\gamma \chi_{c0}(2P)$ & 3$\times 10^{-1}$ /7$\times 10^{-3}$  \\
                  & $\gamma \chi_{c1}(1P)$ & 5$\times 10^{-1}$  & 0.13/0.36 \\
                  & $\gamma \chi_{c2}(1P)$ & 1$\times 10^{-1}$  & 0.075/0.11 \\
                  & $\gamma \chi_{c2}(2P)$ & 4$\times 10^{-2}$   \\
\hline
 $h_c(3P)$        & $\gamma \chi_{c0}(1P)$ & 10 \\
                  & $\gamma \chi_{c0}(2P)$ & 3/2 \\
                  & $\gamma \chi_{c0}(3P)$ & 2$\times 10^{-1}$  \\
                  & $\gamma \chi_{c1}(1P)$ & 5$\times 10^{-1}$  \\
                  & $\gamma \chi_{c1}(2P)$ & 5$\times 10^{-2}$  \\
                  & $\gamma \chi_{c2}(1P)$ & 4$\times 10^{-1}$  \\
                  & $\gamma \chi_{c2}(2P)$ & 1$\times 10^{-3}$  \\
  \hline
  $h_c(4P)$      & $\gamma \chi_{c0}(1P)$ & 14  \\
                 & $\gamma \chi_{c0}(2P)$ & 6/4 \\
                 & $\gamma \chi_{c0}(3P)$ & 2  \\
                 & $\gamma \chi_{c0}(4P)$ & 5$\times 10^{-1}$   \\
                 & $\gamma \chi_{c1}(1P)$ & 5$\times 10^{-1}$   \\
                 & $\gamma \chi_{c1}(2P)$ & 8$\times 10^{-2}$   \\
                 & $\gamma \chi_{c1}(3P)$ & 2$\times 10^{-1}$   \\
                 & $\gamma \chi_{c2}(1P)$ & 1  \\
                 & $\gamma \chi_{c2}(2P)$ & 1$\times 10^{-1}$   \\
                 & $\gamma \chi_{c2}(3P)$ & 9$\times 10^{-3}$   \\
\hline
 $h_c(5P)$       & $\gamma \chi_{c0}(1P)$ & 18   \\
                 & $\gamma \chi_{c0}(2P)$ & 8/6  \\
                 & $\gamma \chi_{c0}(3P)$ & 4   \\
                 & $\gamma \chi_{c0}(4P)$ & 2   \\
                 & $\gamma \chi_{c0}(5P)$ & 1   \\
                 & $\gamma \chi_{c1}(1P)$ & 1   \\
                 & $\gamma \chi_{c1}(2P)$ & 9$\times 10^{-2}$    \\
                 & $\gamma \chi_{c1}(3P)$ & 2$\times 10^{-1}$    \\
                 & $\gamma \chi_{c1}(4P)$ & 2$\times 10^{-3}$    \\
                 & $\gamma \chi_{c2}(1P)$ & 1   \\
                 & $\gamma \chi_{c2}(2P)$ & 3$\times 10^{-1}$    \\
                 & $\gamma \chi_{c2}(3P)$ & 1$\times 10^{-1}$    \\
                 & $\gamma \chi_{c2}(4P)$ & 5$\times 10^{-4}$    \\
\toprule[1.5pt]
\end{tabular}
\caption{The M1 radiative transitions of $^1P_1$ charmonium states. The unit is keV.}\label{M1-3}
\end{table}

\bibliographystyle{apsrev4-1}
\bibliography{Ref}

\end{document}